\documentclass[aps, prb, preprint, amsmath,amssymb,showpacs,superscriptaddress]{revtex4-1}

\usepackage{graphicx}
\usepackage{color}

\makeatletter

\begin{document}

\title{Asymmetric mass  acquisition in LaBi - a new topological semimetal candidate}

\author{Yun Wu}
\affiliation{Division of Materials Science and Engineering, Ames Laboratory, Ames, Iowa 50011, USA}
\affiliation{Department of Physics and Astronomy, Iowa State University, Ames, Iowa 50011, USA}

\author{Tai Kong}
\affiliation{Division of Materials Science and Engineering, Ames Laboratory, Ames, Iowa 50011, USA}
\affiliation{Department of Physics and Astronomy, Iowa State University, Ames, Iowa 50011, USA}

\author{Lin-Lin Wang}
\affiliation{Division of Materials Science and Engineering, Ames Laboratory, Ames, Iowa 50011, USA}

\author{D. D. Johnson}
\affiliation{Division of Materials Science and Engineering, Ames Laboratory, Ames, Iowa 50011, USA}
\affiliation{Department of Physics and Astronomy, Iowa State University, Ames, Iowa 50011, USA}
\affiliation{Department of Materials Science and Engineering, Iowa State University, Ames, Iowa 50011, USA}

\author{Daixiang Mou}

\author{Lunan Huang}

\author{Benjamin Schrunk}
\affiliation{Division of Materials Science and Engineering, Ames Laboratory, Ames, Iowa 50011, USA}
\affiliation{Department of Physics and Astronomy, Iowa State University, Ames, Iowa 50011, USA}

\author{S.~L.~Bud'ko}

\author{P. C. Canfield}
\email[]{canfield@ameslab.gov}

\author{Adam Kaminski}
\email[]{kaminski@ameslab.gov}
\affiliation{Division of Materials Science and Engineering, Ames Laboratory, Ames, Iowa 50011, USA}
\affiliation{Department of Physics and Astronomy, Iowa State University, Ames, Iowa 50011, USA}

\begin{abstract}
We use our high resolution He-lamp based, tunable laser-based ARPES measurements and density functional theory calculations to study the electronic properties of LaBi, a binary system that was proposed to be a member of a new family of topological semimetals. Both bulk and surface bands are present in the spectra. The dispersion of the surface state is highly unusual. It resembles a Dirac cone, but upon closer inspection we can clearly detect an energy gap. The bottom band follows roughly a parabolic dispersion. The dispersion of the top band remains very linear,  ``V" shape like, with the tip approaching very closely to the extrapolated location of Dirac point. Such asymmetric mass acquisition is highly unusual and opens a possibility of a new topological phenomena that has yet to be understood.
\end{abstract}
\date{\today}
\maketitle

The discovery of Quantum Hall Effect \cite{Klitzing1980PRL} introduced the concept of quantum states that can not be classified by spontaneous symmetry breaking, but instead are classified by their topology. Another topological state, Quantum Spin Hall state, has been theoretically predicted and experimentally observed in HgTe quantum wells \cite{Bernevig2006Sci, Konig2007Sci}. This new topological state exists in a system that is insulating in its bulk but topologically conducting on the edge (i. e. 1D equivalent of 2D surface). A Bi$_{0.9}$Sb$_{0.1}$ binary was the first bulk material verified to be a topological insulator by use of angle-resolved photoemission spectroscopy (ARPES) to directly probe the electronic structure \cite{Fu2007PRB, Hsieh2008Nat}. However, its complicated surface states, fairly small bulk band gap and alloying disorder made it hard to be a model system for studying topological quantum phenomena and technological applications. Bi$_2$Se$_3$, Bi$_2$Te$_3$ and Sb$_2$Te$_3$ were theoretically predicted \cite{Zhang2009NatPhys} and experimentally proved to be the second generation topological insulators (or, at least, near insulators) with a single Dirac cone residing at the $\Gamma$ point \cite{Xia2009NatPhys, Chen2009Sci}. The surface Dirac cone states are protected by time reversal symmetry (TRS). Therefore, TRS breaking sources, such as magnetic field or magnetic dopant can modify the massless electrons into finite mass electrons \cite{Chen2010Sci}. 

Discovery of such topologically protected quantum states generated a lot of interest and sparked search for other novel, exotic topological states, such as three-dimensional Dirac semimetals \cite{Wang12PRB, Liu14Sci, Wang13PRB, Neupane14NatCom, Liu14NatMat, Narayanan2015Linear}, type-I  and type-II Weyl semimetals \cite{Huang15NatCom, Weng15PRX, Xu15SciDis, Xu15NatPhys, Xu15SciAdv, Soluyanov2015Type, Sun15Prediction, Chang2016Prediction, Belopolski2015Unoccupied, Huang2016arXiv, Bruno2016Surface, Wang2016Spectroscopic, Wu2016Observation, Wu2015PRL, Borisenko2016Time, XuSY2016Discovery} and line node semimetals \cite{Bian2016Topological, Schoop15arXiv, Wu2016Dirac}. However, no new family of binary topological insulators was reported to date. Recently, simple rock salt rare earth monopnictides LaX (X = N, P, As, Sb, Bi) were predicted to host novel topological states, such as ``linked nodal ring'' in LaN when spin-orbital coupling is neglected \cite{Zeng2015Topological}. When considering the spin-orbital coupling, LaN turns into a three-dimensional Dirac semimetal and the rest of the family turn into topological insulators \cite{Zeng2015Topological}. 

Here, we present the results from our laboratory-based ARPES measurements and density functional theory (DFT) calculations detailing the electronic structure of LaBi.  We observe the coexistence of the bulk and surface states at the $\Gamma$ point from our He lamp and ultrahigh resolution laser based ARPES measurements. The dispersion of the surface state is highly unusual. It resembles a Dirac cone, but upon closer inspection we can clearly detect an energy gap. The bottom  band follows roughly a parabolic dispersion. The top band has an unusual linear ``V" shape dispersion with the tip approaching very closely to the extrapolated location of Dirac point. This is evidence of abnormal, asymmetric mass acquisition by Dirac Fermions. Our data suggests that this compound hosts unusual, yet to be understood topological state.

Single crystals of LaBi were grown using a high-temperature solution growth technique \cite{Canfield92PMPB}. Starting elements (La from Ames Laboratory and Bi from Alfa Aesar, 99.99$\%$ purity) were packed in a frit-disc alumina crucible set (otherwise known as a Canfield Crucible Set or CCS) \cite{Canfield16PhilMag} with a molar ratio of La$\colon$Bi = 30$\colon$70. The crucible with the starting materials were sealed in a silica ampoule under a partial argon atmosphere. The whole ampoule was then heated up to 1200 $^\circ$C, held at 1200~$^\circ$C for 3 hours and slowly cooled to 1000~$^\circ$C over 50--100 hours, at which temperature the solution and the single crystals were quickly separated in a centrifuge. Single crystals of LaBi are cubic in shape with a typical edge length of 0.5 mm. 
ARPES measurements were carried out using Helium discharge lamp (angular and energy resolutions  set at $\sim$ 0.3$^\circ$ and 15 meV, respectively) and tunable, laser based\cite{Jiang14RSI} $\sim$ (0.05$^\circ$ and 1 meV) ARPES spectrometers. Data from the laser-based ARPES system were collected with a tunable photon energy from 5.64 eV to 6.70 eV and the size of the photon beam on the sample was $\sim$30 $\mu$m. Samples were cleaved \textit{in situ} at a base pressure lower than $1 \times 10^{-10}$ Torr. Samples were cleaved at 37K in the He-lamp system and 40K in the laser-based system and were kept at the cleaving temperature throughout the measurements. The cleaved surface is perpendicular to the (100) direction. 
DFT calculations\cite{Hohenberg64PR, Kohn65PR}  have been done in VASP \cite{Kresse96PRB,Kresse96CMS} using PBE \cite{Perdew96PRL} exchange-correlation functional, plane-wave basis set with projected augmented waves \cite{Blochl94PRB} and spin-orbital coupling (SOC) effect included. For bulk band structure of LaBi, we use the conventional tetragonal cell of 4 atoms along (001) direction with a ($10 \times 10 \times 8$) $k$-point mesh. For (001) surface band structure, we use slabs up to 48 atomic layers or 96 atoms with a ($10 \times 10 \times 1$) $k$-point mesh and at least a 12 \text{\AA} vacuum. The kinetic energy cutoff is 165 eV. The convergence with respect to $k$-point mesh was carefully checked, with total energy converged below 1 meV/atom. We use experimental lattice parameters of $a=$6.5799 \text{\AA} with atoms fixed in their bulk positions.

\begin{figure*}[tb]
	\includegraphics[width=6in]{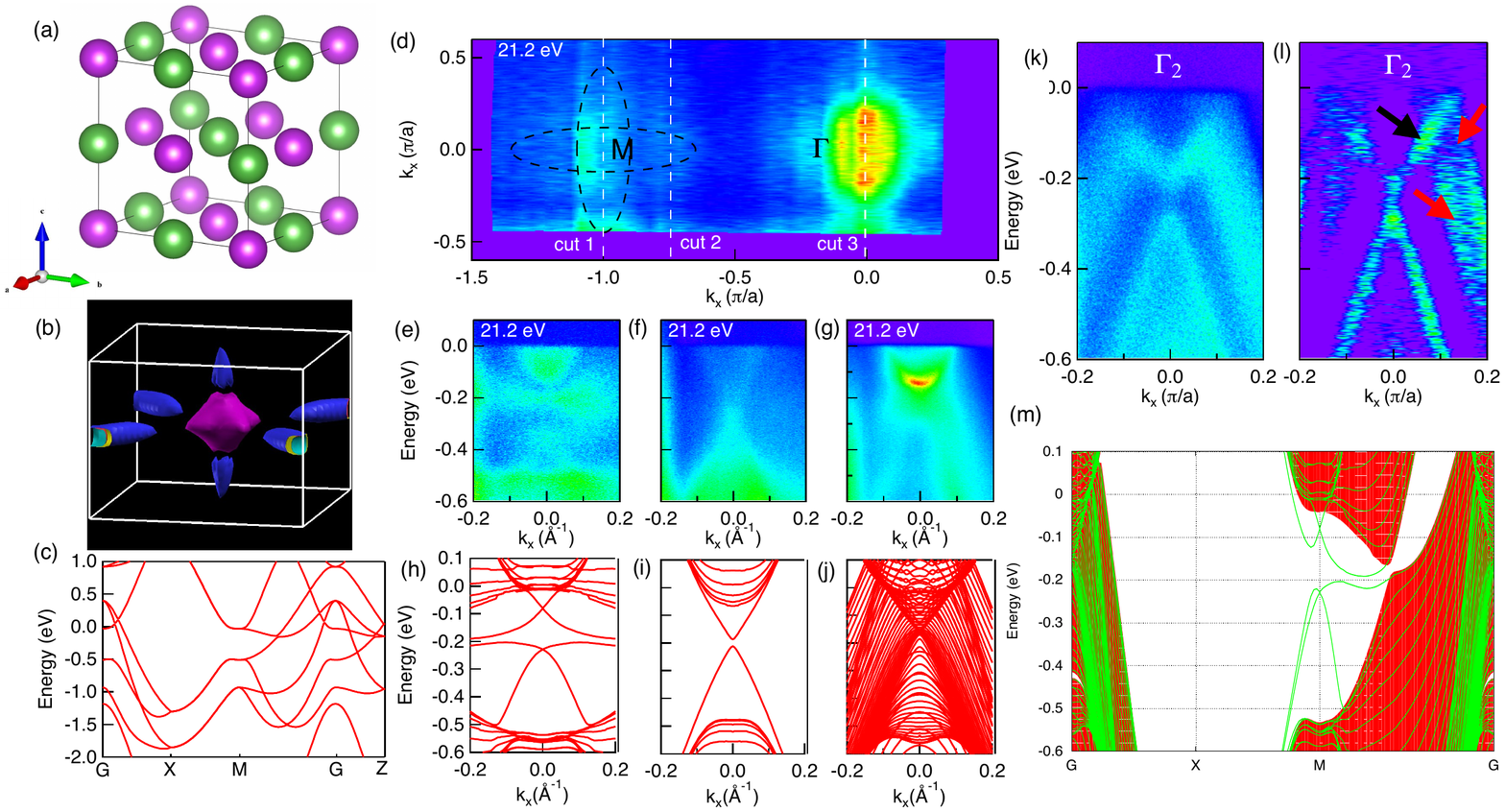}%
	\caption{(color online) Calculated and experimental Fermi surface  (FS) and band dispersion of LaBi measured at $T=$37 K and photon energy of 21.2 eV. 
	(a) Crystal structure (La: purple spheres, Bi: green spheres) of LaBi.
	(b) Brillouin zone (BZ) and DFT-calculated 3D bulk FS of LaBi.
	(c) Calculated bulk dispersion along main symmetry directions.
	(d) FS plot of ARPES intensity integrated within 10 meV of the chemical potential along $\Gamma-M$.
	(e)--(g) ARPES intensity along cuts 1 -- 3 marked by white dashed lines in (b).
	(h)--(j) Surface-band dispersion calculated for a 48-layer slab along cuts 1 -- 3 in (b) .
	(k) Measured dispersion along $\Gamma$ cut in second BZ.
	(l) Second derivative of data in (k). Black and arrows point to electron and hole bands respectively.
	(m) Projection of 3D bulk dispersion in red with overlapped green surface bands calcukated for a 48-layer slab.
	\label{fig:Fig1}}
\end{figure*}

The crystal structure, calculated 3D Fermi Surface (FS) and band dispersion along key directions in the Brillouin Zone (BZ) for LaBi are shown in Fig.\ref{fig:Fig1}(a-c). Panel (d) shows the ARPES intensity measured at the chemical potential using He-I line (21.2 eV) at $T=37$ K. The data was integrated within 10 meV to improve statistics. High intensity areas mark the contours of the FS sheets. The FS consists of one electron and two hole pockets at the $\Gamma$ point and two elliptical electron pockets at the $M$ point (black dashed lines are guide to the eye). The FS resembles the calculated bulk-band FS from DFT as shown in Fig.\ref{fig:Fig1}b. 
Panels (e)--(g) show the band dispersion measured using ARPES along cuts 1--3 (marked in (b) as white dashed lines) in Fig.\ref{fig:Fig1}d. Panels (h)--(j) show the corresponding surface band calculations with 48-layer slab along those same cuts shown in panel (e)--(g). 
In panel (e), we can see two electron pockets at M point with the smaller one being enclosed by the bigger one, which agrees with the calculations shown in panel (c). Panel (f) shows the band dispersion along the cut 2 at the crossing point of the $d-p$ orbital mixing. This feature may look  like a Dirac cone, except that the calculation shows a possible gap separating the top and bottom bands.  Our DFT calculations results are similar with the results in Ref.\cite{Zeng2015Topological} in which topological surface state was predicted to reside in the $d-p$ band inversion regime. However, due to limited resolution and limited tunability of the photon energy in the He-lamp ARPES system, we cannot verified its surface origin by probing its out of plane momentum dispersion in the proximity to the M point. At the $\Gamma$ point (panel (g)), an electron pocket is clearly seen. However, no details can be resolved at higher binding energies. 
Panel (j) show the calculated surface-band dispersion along the same cut as in panel (g), which very roughly resembles main features measured by ARPES results. The electron pocket and two hole pockets are clearly observed in the second BZ, as shown in Fig.\ref{fig:Fig1}k and its second derivative in panel (l). 
The band dispersion of the surface state at $\Gamma$ is more complicated, because there is no gap in the projected 3D bulk dispersion, as shown in Fig.\ref{fig:Fig1}m. This means that signal from both bulk and surface states will  contribute to photoelectron intensity. 

\begin{figure}[bt]
	\includegraphics[width=6in]{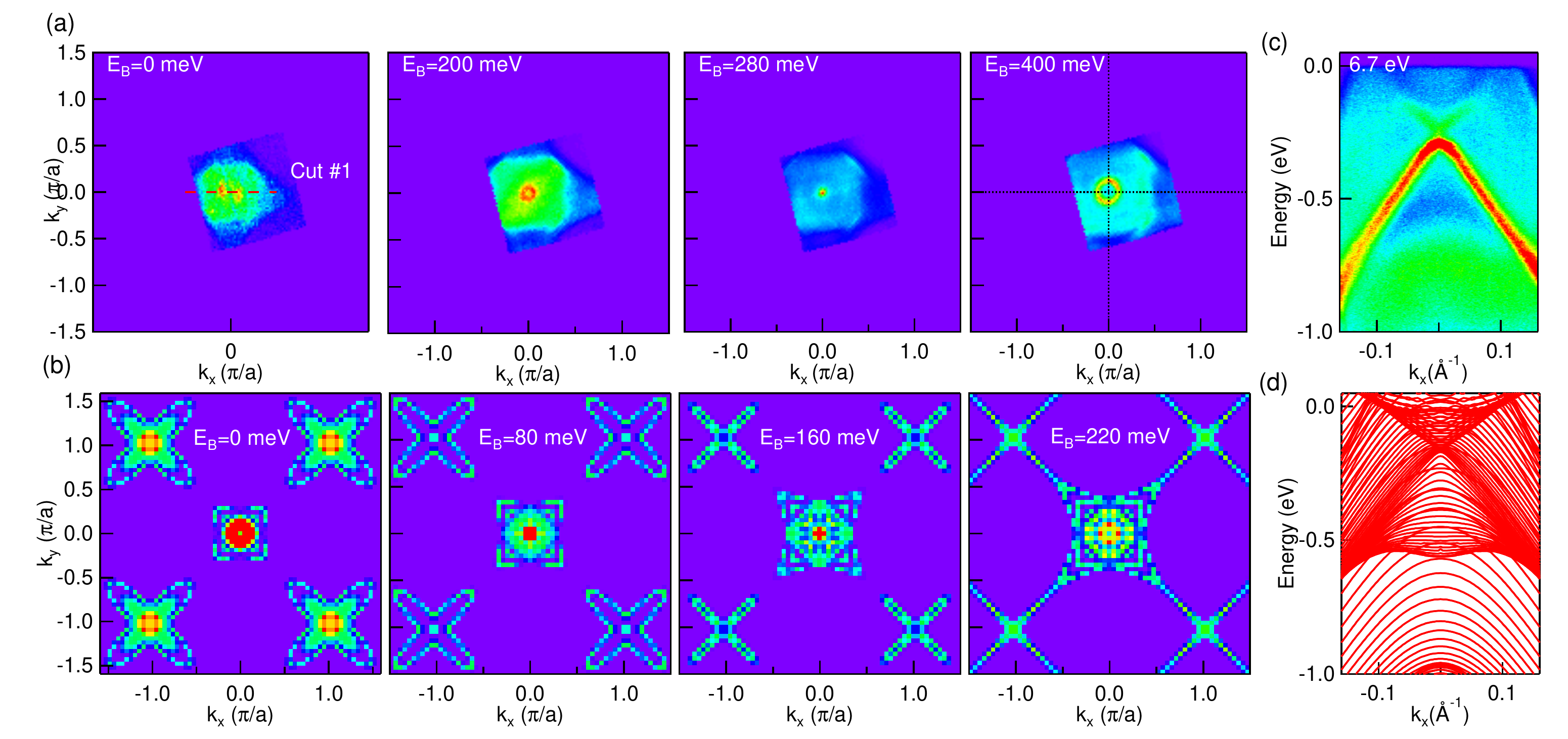}%
	\caption{(color online) Fermi surface and band dispersion in the proximity of the $\Gamma$ point measured at $T=$ 40 K and photon energy of 6.70 eV. 
	(a) Constant energy contour plots of ARPES intensity integrated within 10 meV at the binding energy of 0, 200, 280 and 400 meV.
	(b) Constant energy contour plots of DFT surface-band calculation at the binding energy of 0, 80, 160 and 220 meV with 16-layer slab.
	(c) Band dispersion along cut 1 marked in panel (a).
	(d) Calculated surface-band dispersion along $\Gamma-X$ in panel a with 48-layer slab.
	\label{fig:Fig2}}
\end{figure}

To reveal the details of these states at $\Gamma$ we used vacuum ultraviolet laser ARPES spectrometer. The low photon energy combined with small beam spot and ultrahigh resolution allows us to gain more information about these features. Figure 2 shows the constant energy contours and data along high symmetry cut along with results of DFT surface-band calculations using a slab method. Panel a shows the constant energy contours measured at 40 K and photon energy of 6.7 eV. 
The constant energy contour at the Fermi level shows rather blurred features dominated mostly by bulk bands. 
At the binding energy of 200 meV, a circular energy contour can be clearly observed, surrounded by square shape bulk band intensities. Further moving down to 280 meV below the Fermi level, the circle shrinks to a dot of intensity. At binding energy of 400 meV, the dot expands to almost perfect circle. Panel (b) shows the constant energy contours from DFT calculations with 16-layer slab, which also shows the evolution of the Dirac cone-like feature from a circular contour to a single Dirac point and further to a circular contour, which is not very easily resolved due to contribution of the bulk band projection, but has overall shape consistent with the data. The surface Dirac cone-like band dispersion can be better visualized in band dispersion data (panel c) along cut 1 in Fig.\ref{fig:Fig2}a. The bulk conduction band crosses the Fermi level and the top of the bulk valence band is visible in panel c. The conduction and valence band appear to be connected by a surface state that forms Dirac-like cone. Panel d shows the calculated surface state with 48-layer slab, which demonstrates that the surface state is buried in the bulk state projection. This is consistent with the data shown in panel a and it is also consistent with previously reported results \cite{Zeng2015Topological}.

\begin{figure}[bt]
	\includegraphics[width=6in]{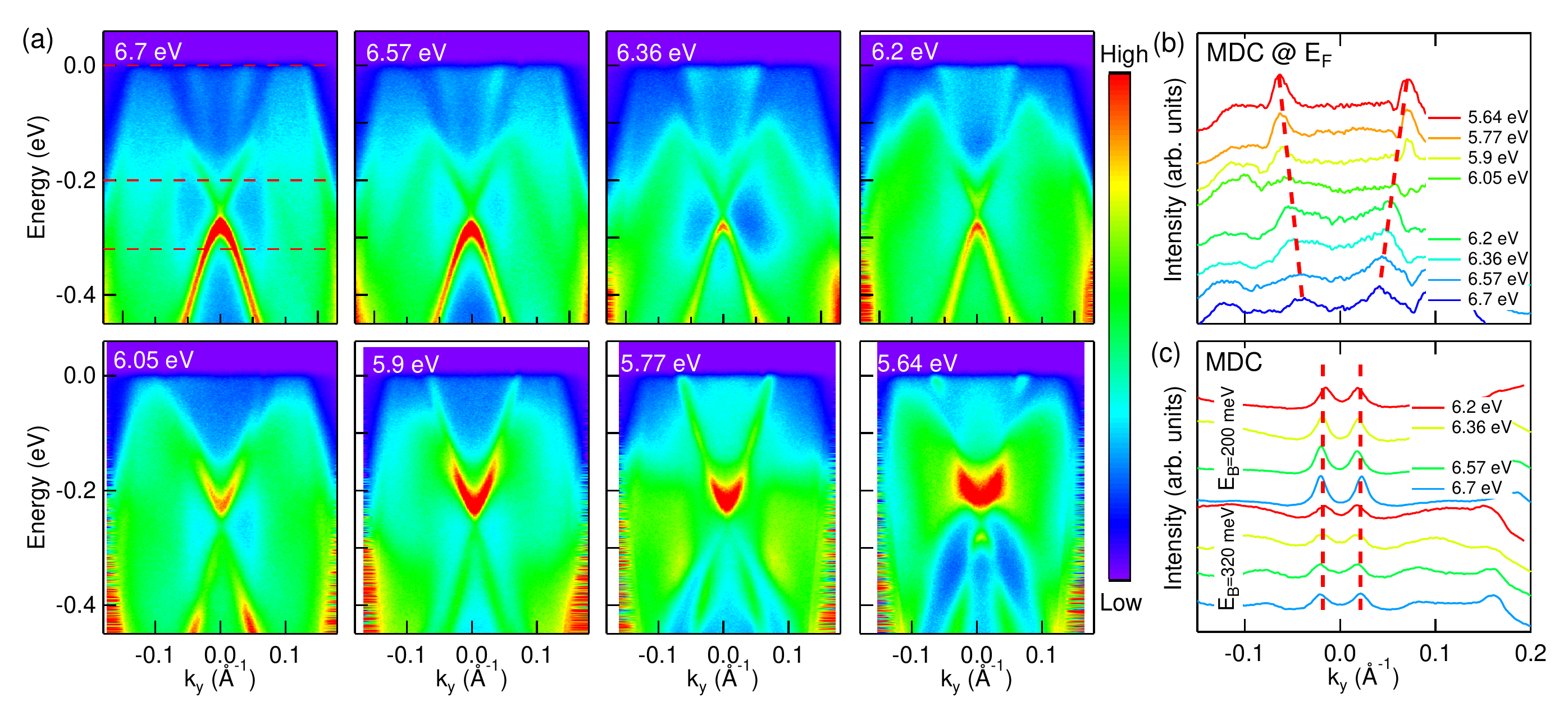}%
	\caption{(color online) Band dispersion measured at $T=$ 40 K using several photon energies. 
	(a) Band dispersion along cut 1 in Fig.\ref{fig:Fig2}a using photon energy of 6.70, 6.57, 6.36, 6.20, 6.05, 5.90, 5.77, and 5.64 eV.
	(b) Momentum dispersion curves at the chemical potential for data in panel (a).
	(c) Momentum dispersion curves at the binding energy of 200 meV and 320 meV for data in panel (a).
	\label{fig:Fig3}}
\end{figure}

We utilize  photon energy dependent ARPES data to distinguish between bulk and surface states as shown in Fig.\ref{fig:Fig3}. A single Dirac-like dispersion is present at higher photon energies (top row of data in panel (a)) with no obvious change in shape. However, the size of the conduction electron pocket and intensity of bulk hole band change drastically especially for lower photon energies and overshadow the surface state due to different matrix elements (bottom row of panel (a)). In order to qualitatively determine the change in the size of the conduction electron pocket as a function of k$_z$ momentum, we have plotted the momentum dispersion curves  (MDCs) at the Fermi level in panel (b), which clearly shows an increase of the electron pocket size with decreasing incident photon energies. For the four highest photon energy we plot the MDC's at binding energies of 200 meV (top part of Dirac cone-like feature) and 320 meV (bottom part of Dirac cone-like feature). Constant separation between the MDC peaks demonstrates surface origin or quasi two dimensionality of this feature. 

\begin{figure}[tb]
	\includegraphics[width=6in]{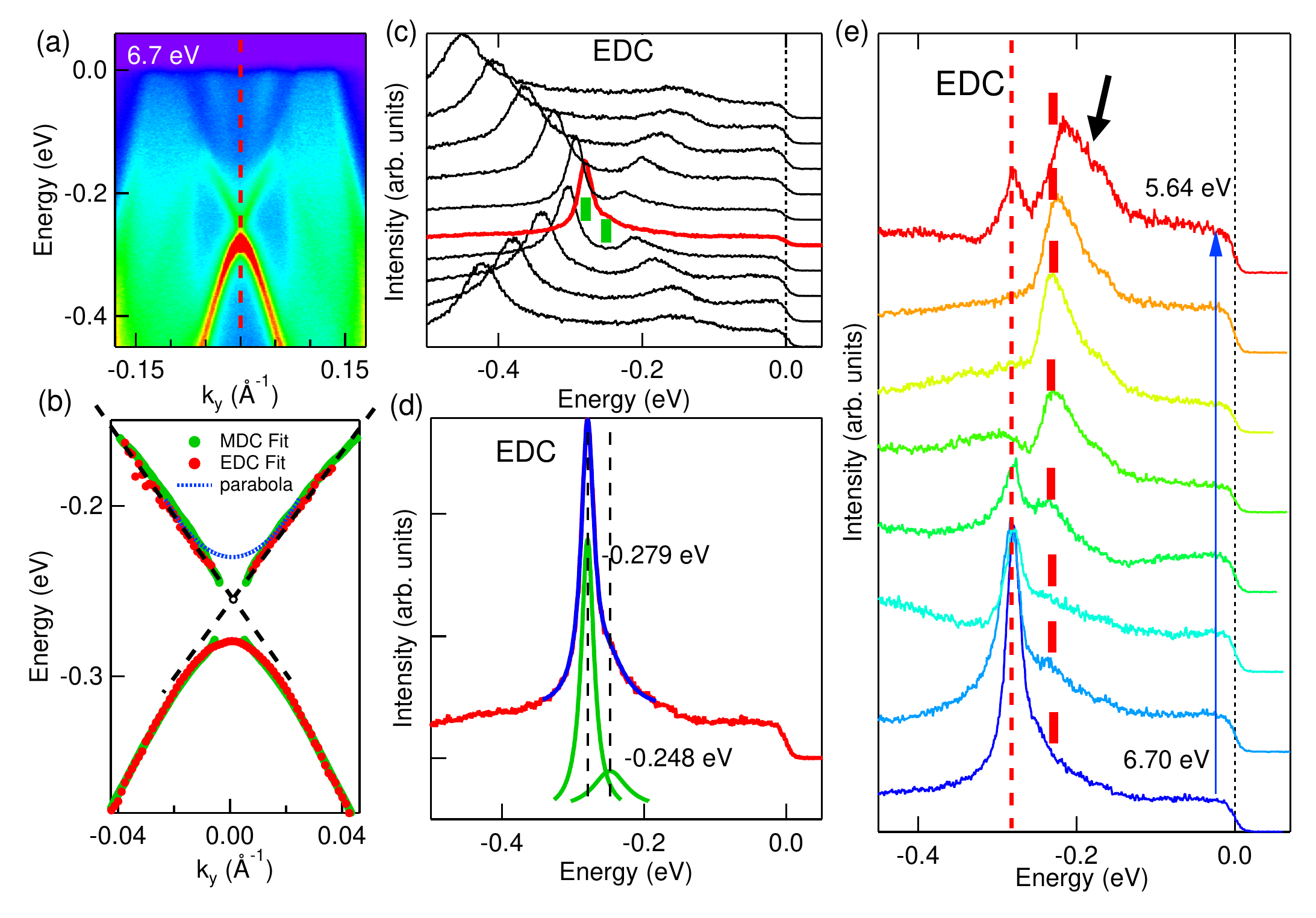}%
	\caption{(color online) Band dispersion and EDCs measured at $T=$40K and photon energy of 6.7 eV.
	(a) Band dispersion measured at along symmetry direction at $\Gamma$. 
	(b) Band dispersion extracted by MDC (green) and EDC (red) fits. The black dashed lines are extension of the top Dirac like bands. Blue dotted line marks the dispersion of the bottom band reflected about the energy of Dirac point - i. e. show the expected dispersion of the upper band for the case of symmetrical mass acquisition.
	(c) set of equally spaced EDC's corresponding to the data in (a). Red curve is measured at $\Gamma$ and it reveals presence of energy gap separating upper and lower branches marked by bars.
	(d) Single EDC corresponding to the data in (c). The green curves are two Lorentzian curves fitted to the EDC and the blue curve is the composite of the two green curves. The black dashed lines mark the location of the peak positions.
	(e) EDC curves at $\Gamma$ for measured photon energies. Red bar marks location of surface state peak. Black arrow marks intensity due to bulk band that increases at lower photon energies.
	\label{fig:Fig4}}
\end{figure}

The key question raised by these data is whether or not this actually is a relativistic, Dirac dispersion with no energy gap and apparent degeneracy of electronic states at the Dirac point. To examine this we use EDC's and look for the presence of an energy gap. The band dispersion along $\Gamma$ cut measured with 6.7 eV photons is shown in Fig. \ref{fig:Fig4}(a). In Fig. \ref{fig:Fig4}(b), we show the band dispersion extracted from MDC peaks (green lines) and EDC peaks (red lines). 
The lower band has a parabolic dispersion that can only occur if an energy gap is present. To verify this, we have plotted a set of EDC's equally spaced in the momentum in Fig. \ref{fig:Fig4}(c). The EDC at the suspected location of the Dirac point shows a peak that originates from bottom part of the dispersion and a distinct shoulder at lower binding energy that originates from the upper band. The two peaks fitted to EDC at the $\Gamma$ point are shown in Fig. \ref{fig:Fig4}(d). We also verified that at other photon energies, where the matrix elements weaken the intensity of the bottom band, clear dip is observed in EDCs  at the energy that would correspond to the Dirac point - an evidence that a gap is present instead \ref{fig:Fig4}(e). Note that at very low photon energies the bulk intensity ovelaps and moves the apparent location of the upper peak to even lower binding energies. This addition intensity is indicated by black arrow in top curve of panel (e). These data confirm the presence of an energy gap separating the two bands and it demonstrates that Dirac fermions acquire mass and energy gap. 

This is not a case of a trivial band gap. While the bottom part of the band is parabolic, the top part is remarkably linear with a pronounced cusp pointing towards the bottom band. Usually, when Dirac fermions acquire mass, the upper and lower bands should develop similar parabolic features with degree of symmetry about the energy of the Dirac point. 
The experimental data is very different, as the upper band remains linear and cuspy almost to the Dirac point energy. To better illustrate this we marked the expected dispersion of the upper band for the case of symmetric mass acquisition by blue dots in \ref{fig:Fig4}(b). Such asymmetric acquisition of mass was not predicted by theory to the best of our knowledge and further theoretical efforts are needed to explain this highly unusual behavior.

In summary, we studied the electronic properties of newly proposed topological semimetal LaBi. The dispersion of the surface state resembles a Dirac cone, but upon closer inspection we can detect an energy gap. The bottom  band follows roughly a parabolic dispersion. The top band has an unusually linear, ``V'' shape dispersion with the tip approaching very closely to the bottom band. Such abnormal, asymmetric mass acquisition by Dirac Fermions suggests that this compound likely hosts unusual, yet to be understood topological state.

\section*{Acknowledgements}
We acknowledge very useful discussions with Yuan-Ming Lu. This work was supported by the U.S. Department of Energy, Office of Science, Basic Energy Sciences, Materials Science and Engineering Division. Ames Laboratory is operated for the U.S. Department of Energy by Iowa State University under contract No. DE-AC02-07CH11358. L. H. was supported by CEM, an NSF MRSEC, under grant DMR-1420451.  

\section*{Corresponding Author}
Correspondence to: Adam Kaminski, email: kaminski@ameslab.gov; Paul C. Canfield, email: canfield@ameslab.gov.

\section*{References}
\bibliography{LaBi}
\newpage

\end{document}